\documentclass[a4paper]{aa}
\usepackage{txfonts,psfig,multirow}
\usepackage{bm}
\begin{document}

\def\''{\lq\lq}
\def\rot{\mathop{\rm rot}\nolimits}
\def\div{\mathop{\rm div}\nolimits} 
\renewcommand{\vec}[1]{\mbox{\boldmath $#1$}}
 
\def\solar{\ifmode_{\mathord\odot}\else$_{\mathord\odot}$\fi} %Sun% 
\def\gsim{\lower.4ex\hbox{$\;\buildrel >\over{\scriptstyle\sim}\;$}} 
\def\lsim{\lower.4ex\hbox{$\;\buildrel <\over{\scriptstyle\sim}\;$}} 
\def\~{$\sim$} 
\def\x{\times} 
 
\def\alf{$\alpha$}
\def\L{$\Lambda$}
\def\nT{$\nu_{\rm T}$\ }
\def\mT{$\mu_{\rm T}$\ }
\def\cT{$\chi_{\rm T}$\ }

\def\apj{{ApJ}}       
\def\apjs{{ Ap. J. Suppl.}} 
\def\apjl{{ Ap. J. Letters}} 
\def\pasp{{ Pub. A.S.P.}} 
\def\mn{{MNRAS}} 
\def\aa{{A\&A}} 
\def\aasup{{ Astr. Ap. Suppl.}} 
\def\baas{{ Bull. A.A.S.}\ } 
\def\csss{{Cool Stars, Stellar Systems, and the Sun}}
\def\an{{Astron. Nachr.}}
\def\sp{{Sol. Phys.}}   
\def\gafd{{Geophys. Astrophys. Fluid Dyn.}} 
\def\acta{{Acta Astron.}}
\def\jfm{{J. Fluid Mech.}}
\def\cpc{{Comput. Phys. Commun.}}
\def\Om{{\it \Omega}}
\def\Omst{\Omega^*}
\def\R{R\"udiger}

\def\qq{\qquad\qquad}                      % horizontaler Platz
\def\qqq{\qquad\qquad\qquad}               % horizontaler Platz
\def\q{\qquad}
\def\bib{\item{}}
\def\top{\item}
\def\toptop{\itemitem}
\def\start{\begin{itemize}}
\def\stop{\end{itemize}}
\def\beg{\begin{equation}}
\def\ende{\end{equation}}
\def\la{\langle}
\def\ra{\rangle}

\title{The  eddy  heat-flux in rotating turbulent convection }
\author{G. \R\ \and P. Egorov \and L.L. Kitchatinov \and M. K\"uker}
\offprints{gruediger@aip.de}
\institute{Astrophysical Institute Potsdam,  
An der Sternwarte 16, D-14482 Potsdam, Germany }

\date{\today}

\abstract{
The three components of the heat-flux vector $\vec{F}=\rho C_{\rm p} \langle \vec{u}'
T'\rangle$ are numerically computed for a stratified rotating turbulent
convection using the NIRVANA code in a flat box. The latitudinal component
$F_\theta$ proves to be  negative (positive) in the northern (southern) hemisphere so that
the heat always flows towards the poles. As a surprise, the radial heat-flux $F_r$ peaks at the equator rather than at the
poles (Taylor numbers  $O(10^6)$). The same behavior is observed for   the radial turbulence intensity $\langle u_r'^2\rangle$ which for {\em free} turbulence is also believed to peak at the poles (see Eq. (\ref{upolequat}) below). As
we can show, however, the consequences of this  unexpected result
(also obtained by K\"apyl\"a, Korpi \& Tuominen 2004) for the theory of
differential rotation are small as mainly the $F_\theta$ is responsible to solve
the `Taylor number puzzle'.\\
In all our simulations the azimuthal component $F_\phi$ proves to be negative so that the rotating turbulence produces  an   westwards directed azimuthal heat-flux which should be  observable. Fluctuations with higher temperature are expected to  be anticorrelated with their own angular velocity fluctuations. 
We find  this rotation-induced result as understandable as the $F_\phi$ is closely
related to the radial $\Lambda$-effect which is known to be also negative in
stratified and rapidly rotating convection zones.
\keywords{Convection, Turbulence, Sun: granulation, Sun: rotation }}
\authorrunning{G. \R\  et al.}
\titlerunning{Heat transport in rotating turbulence}
\maketitle

%%%%%%%%%%%%%%%%%%%%%%%%%%%%%%%%%%%%%%%%%%%%%%%%%%%%%%%%%%%%%%%%%%%%%%
\section{Motivation}\label{s1}
There are more and more data describing the surface rotation law of stars, the
majority of which complies both with the sign and amplitude of the known equatorial
acceleration of the Sun. In order to compare the various observational results a
relation
\beg
\Om_{\rm eq} - \Om_{\rm pole} \propto \Om_{\rm eq}^{n''}
\label{Omeq}
\ende
has been introduced with $n''$ representing the $\Om$-dependence of the
equator-pole difference of the surface angular velocity. It proved to be weak in
the first papers in this research ($n''\simeq 0.15$, Hall 1991) while in recent
studies values of 0.58 (Messina \& Guinan 2003) and 0.66 (Reiners \& Schmitt
2003) have been found. A rotational influence upon the rotation law is
clearly existing but   it seems not to be too strong.

To   find the
influence of the global rotation on the turbulent convection in the stellar
convection zones is thus the basic problem of the theory of differential rotation. The resulting flow
pattern (`rotating turbulent convection') simultaneously transports both the heat and the angular momentum. The resulting 
 angular momentum transport has been described
with the  $\Lambda$-effect whose properties easily can be
reproduced with box simulations (see the references in R\"udiger \& Hollerbach 2004). In the present paper it is the eddy heat-flux in rotating
turbulent convection which shall be  rediscussed here by means of new box
simulations.

For the usual Boussinesq relation
\beg
\vec{F}=-\rho T \chi \nabla S
\label{0}
\ende
between the eddy heat-flux and the entropy gradient Weiss (1965) and Durney \& Roxburgh (1971)
 parameterized the influence of a global rotation by
\beg
\chi = \chi_0\left(1+\varepsilon \cos^2\theta\right)
\label{01}
\ende
with $\varepsilon$ representing the rotational influence. The eddy conductivity is
assumed to involve the main influence of the turbulent flow pattern which due to
the basic rotation strongly depends on the colatitude $\theta$. The ansatz
(\ref{01}) leads to a $\theta$-dependence of all thermodynamic
quantities with a temperature difference of pole and equator so that  a meridional
flow is  the immediate consequence. There was no possibility at that time for a
detailed theory of the connection of $\chi_0$ and $\varepsilon$ with the
characteristic parameters of the turbulence (see also Belvedere, Patern\`o \& Stix 1980).

In the magnetohydrodynamics of the Sun meridional flows are playing a more and
more important role. The theory of the differential rotation in the convection
zone and the tachocline as well as the theory of the solar dynamo are important 
examples. We shall discuss details of the thermodynamics of the rotation theory
 in this paper. We shall derive the ideas for the structure of the
eddy-conductivity heat tensor (next Section). In the following Sections the
results will be compared with the results of
nonlinear simulations with the finite-difference code NIRVANA.

%%%%%%%%%%%%%%%%%%%%%%%%%%%%%%%%%%%%%%%%%%%%%%%%%%%%%%%%%%%%%%%%%%%%%%%% 
\section{Quasilinear theory}\label{s2}
In rotating turbulent fluids the relation between the turbulent heat-flux
$\vec{F}=\rho C_{\rm p} \langle  \vec{u}' T'\rangle$ and the superadiabatic temperature
gradient
\beg
\vec{\beta}= \frac{\vec{g}}{C_{\rm p}}-\nabla T
\label{1}
\ende
($\vec{g}$ gravity) is a tensorial one, i.e.
\beg
F_i=\rho C_{\rm p} \chi_{ij} \beta_j
\label{2}
\ende
with $\vec{\beta}=\nabla T^{\rm ad} - \nabla T=-(T/C_{\rm p})\nabla S$ where $S$
is the
gas entropy. In the simplest case known from the mixing-length theory it is
$\chi_{ij}=\chi_{\rm T} \delta_{ij}$ so that Eq.~(\ref{0}) results (instead of $\vec{F}=-\rho C_{\rm p} \chi_{\rm T} \nabla T$
for incompressible fluids).
%%%%%%%%%%%%%%%%%%%%%%%%%%%%%%%%%%%%%%%%%%%%%%%%%%%%%%%%%%%%%%%%%%%%%%%%%%%%%
\subsection{Isotropic turbulence subject to rotation}\label{s21}
For rotating fluids without any other preferred direction than the rotation vector
$\vec{\Om}$ it is simply
\beg
\chi_{ij}=\chi_{\rm T} \delta_{ij} + \chi_\parallel \Om_i \Om_j +
\tilde \chi \epsilon_{ipj} \Om_{p}
\label{3}
\ende
(Kitchatinov, Pipin \& R\"udiger 1994). The poloidal components of (\ref{3}) in
spherical coordinates are
\begin{eqnarray}
\lefteqn{\chi_{rr}=\chi_{\rm T}+\cos^2\theta \chi_\parallel \Om^2,}\nonumber\\
\lefteqn{\chi_{\theta r}= - \sin\theta \cos\theta \chi_\parallel \Om^2.}
\label{5}
\end{eqnarray}
If the averages are taken over the horizontal plane,
$\vec{\beta}$
has only a radial positive component which is positive in the convectively unstable zones.

With (\ref{5}) the heat-transport in rotating but otherwise isotropic
turbulence is rather simple. If the latitudinal heat-transport goes to the poles
(equator) then the radial heat-transport at the poles is always stronger (weaker) than
at the equator. In the first case the $\chi_\parallel$ is positive and in the
second case it is negative. 

The coefficients $\chi_{\rm T}$ and $\chi_\parallel$
have been computed by Kitchatinov, Pipin \& R\"udiger (1994, their Fig.~1). They
both are positive. The $\chi_{\rm T}$ vanishes like $O(\Om^{-1})$ for fast
rotation while the $\chi_\parallel$ vanishes like $O(\Om^{-3})$. Hence, for
isotropic turbulence under the influence of a global rotation, the radial eddy
heat-flux peaks at the poles, and the latitudinal  heat-flux goes to the poles.
%%%%%%%%%%%%%%%%%%%%%%%%%%%%%%%%%%%%%%%%%%%%%%%%%%%%%%%%%%%%%%%%%%%%%%%%%5
\subsection{Anisotropic turbulence subject to rotation}\label{s22}

The situation is more complicated if the turbulence-field without rotation is
already anisotropic. The anisotropy (unit) vector may be $\vec{G}$. 
In this case extra terms appear in (\ref{3}) which in
the simplest case run with the second order in $\vec{G}$, i.e.
\beg
\chi_{ij}=\dots + \chi_1(\vec{G}\vec{\Om})^2 \delta_{ij} +
\chi_2(\vec{G}\vec{\Om}) (G_i \Om_j + G_j \Om_i).
\label{5.1}
\ende
These terms provide the new contributions 
\begin{eqnarray}
\lefteqn{\chi_{rr}=\dots + (\chi_1+2\chi_2) \cos^2\theta \Om^2}\nonumber\\
\lefteqn{\chi_{\theta r}= \dots - \chi_2\sin\theta \cos\theta \Om^2.}
\label{5.2}
\end{eqnarray}
to the poloidal components (\ref{5}) of the
heat-conductivity tensor.
Hence,  now  
\begin{eqnarray}
\lefteqn{\chi_{rr}=\chi_{\rm T}+ (\hat \chi + \chi_\parallel) \Om^2
\cos^2\theta,}\nonumber\\
\lefteqn{\chi_{\theta r}= - \chi_\parallel \Om^2 \sin\theta\cos\theta,}
\label{5.3}
\end{eqnarray}
where some of the notations have slightly been changed. The sign of the unknown
quantity $\hat \chi$ may fix the latitudinal  profile of the radial
heat-flux.

In the following we shall  compute (\ref{5.3}) for various
rotation rates by means of numerical box simulations. We shall find, indeed,
that the pole-equator difference of the radial heat-flux is a sensitive function
of rotation, stratification and/or boundary conditions. This is not true,
however, for the latitudinal heat-flux which will prove to be towards the poles in
all our simulations.

%%%%%%%%%%%%%%%%%%%%%%%%%%%%%%%%%%%%%%%%%%%%%%%%%%%%%%%%%%%%%%%%%%%%%%%%%%%%%%%%%%%%%%%%%%%%
\subsection{Rotating free turbulence}\label{s23}
There is a close relation of the heat-flux tensor and the one-point correlation
tensor
\beg
Q_{ij}=\langle u_i'(\vec{x},t) u_j'(\vec{x},t)\rangle .
\label{qij}
\ende
In order to demonstrate this basic outline we start from the quasilinear relation
\begin{eqnarray}
\lefteqn{\chi_{ij}=\int \frac{\hat Q_{ij}(\vec{k},\omega)}{-{\rm i}\omega+\chi k^2} {\rm
d}\vec{k} {\rm d}\omega=}\nonumber\\
&& \quad =\int \frac{\chi k^2 \hat Q_{ij} {\rm d}\vec{k} {\rm d}
\omega}{\omega^2+\chi^2 k^4}+
 {\rm i} \int \frac{\omega \hat Q_{ij} {\rm
d}\vec{k} {\rm d}\omega}{\omega^2+\chi^2 k^4}
\label{100}
\end{eqnarray}
between the spectral tensor $\hat Q_{ij}$ of the turbulence and the heat-conductivity
tensor $\chi_{ij}$ (e.g. R\"udiger 1989). 
The last term in (\ref{100})  provides an antisymmetric part of the
$\chi_{ij}$-tensor which is still ignored.

In the remaining term for $\chi \to 0$ a Dirac $\delta$-function is involved so
that in this approximation
\beg
\chi_{ij}= \pi \int \hat Q_{ij}(k,0) {\rm d} k\equiv \frac{1}{2} \int Q_{ij} (0,\tau)
{\rm d}\tau.
\label{11}
\ende
If this relation is applied  and if the $\tau$-integral
is approximated by $\tau_{\rm corr}$ then
\beg
\chi_{ij}=\frac{1}{2} \tau_{\rm corr} Q_{ij}.
\label{chiij2}
\ende
It  makes sense, therefore,  to discuss briefly the one-point correlation
tensor $Q_{ij}$ under the influence of rotation. In the Appendix the over-all
structure of this tensor is given within the $\tau$-approximation which is
adopted in the following. Two properties
are  particularly stressed here: i) the anisotropy of the rotating turbulence at the poles and ii) the
pole-equator difference of the radial turbulence intensity $\langle
u_r'^2\rangle$.

At the poles we have $\langle u_\theta'^2\rangle=\langle u_\phi'^2\rangle$. For
the characterization of the anisotropy it is thus enough to write
\begin{eqnarray}
\lefteqn{\langle u_r'^2\rangle -\langle u_\theta'^2\rangle - \langle
u_\phi'^2\rangle= \langle u_r'^2\rangle - 2\langle u_\phi'^2\rangle=}\nonumber\\
&& =
\bigg(\phi_\parallel-\phi+a(\phi_1+\phi_2+2\phi_3-\phi')\bigg)\langle
u_r^{(0)2}\rangle 
\label{ur}
\end{eqnarray}
with $a$ after Eq.~(\ref{a4}). They will be considered in this Section
exclusively. The azimuthal heat-flux is discussed in Section~\ref{s6}. With the relations (\ref{a13}) it follows for fast rotation that
\beg
\langle u_r'^2\rangle-\langle u_\theta'^2\rangle-\langle u_\phi'^2\rangle\simeq
\frac{3\pi}{8}\frac{a}{\Omst} \langle u_r^{(0)2}\rangle .
\label{urphitheta}
\ende
Here 
\beg
\Omst=2 \tau_{\rm corr} \Om
\ende
 is the Coriolis number (or inverse Rossby number) of the turbulence. Hence, for isotropic  turbulence ($a=0$) 
\beg
\langle u_r'^2\rangle \simeq \langle u_\phi'^2\rangle + \langle
u_\theta'^2\rangle
\label{urfithe}.
\ende
At the poles the (fast) rotation  of an originally isotropic turbulence results in a dominance of the
vertical turbulence components. The effect is reduced  for radial-type turbulences
($a<0$) and it is enhanced for horizontal-type turbulences ($a>0$).

\begin{figure}
  \psfig{figure=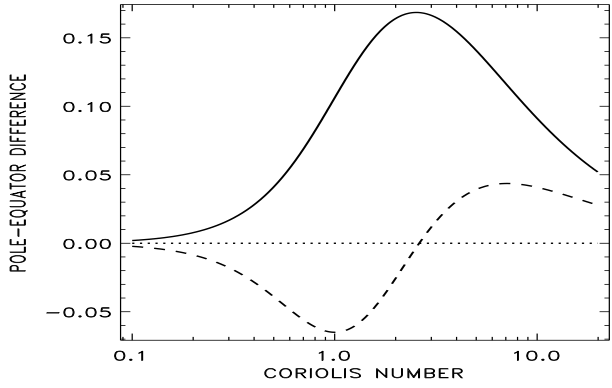,height=5cm,width=8cm}
  \caption{Contributions of the isotropic (full line) and anisotropic
  (dashed line) parts of turbulence into the normalized pole-equator
  difference (\ref{upolequat}) of the radial turbulence intensity. The full and dashed lines are the functions
  $\phi_\|$ and $\phi'_\| + \phi_1$.}
  \label{fa1}
\end{figure}

Canuto, Minotti \& Schilling (1994) demonstrated for
their rather general model how the global rotation  influences
the formation  of anisotropy between the  components of the
turbulence intensity. If in the equatorial plane without rotation both the
turbulent intensities $\langle u_r'^2\rangle$ and $\langle u_\phi'^2\rangle$
strongly differ then the rotation is smoothing the differences and may even
completely suppress them (R\"udiger, Tsch\"ape \& Kitchatinov 2002).
For faster rotation there is thus a
clear tendency for a  {\em return-to-isotropy}. A similar
(even crossover) behavior for anisotropic turbulence  has
also been found for his rotating convection-turbulence  by
Chan (2001).  

For the radial turbulence intensity one finds the relation
\beg
\langle u_r'^2\rangle \bigg|_{\rm pole} - \langle u_r'^2\rangle \bigg|_{\rm
eq} = \left(\phi_\parallel + a\left(\phi'_\parallel + \phi_1\right)\right)
\langle u_r^{(0)2}\rangle .
\label{upolequat}
\ende
The contributions of the isotropic and anisotropic parts of turbulence
to the pole-equator difference (\ref{upolequat}) are shown in Fig.~\ref{fa1}
as functions of the Coriolis number. For radial-dominated free turbulence
($a<0$) the expression (\ref{upolequat}) is always positive.
Therefore, $\langle u_r'^2\rangle$ for rotating turbulence should be always greater at
the poles rather than at the equator.
%%%%%%%%%%%%%%%%%%%%%%%%%%%%%%%%%%%%%%%%%%%%%%%%%%%%%%%%%%%%%%%%%%%%%%%%% 

%%%%%%%%%%%%%%%%%%%%%%%%%%%%%%%%%%%%%%%%%%%%%%%%%%%%%%%%%%%%%%%%%%%%%%%
\section{Basic equations, the model}

3D numerical simulations of compressible, thermal convection under 
the influence of rotation are made with
the finite-difference, fractional-step code NIRVANA (Ziegler 1998, 
 1999) in a small rectangular box defined 
on a Cartesian grid. It can be considered
as a small piece of a spherical shell. The domain is placed tangentially at  different
colatitudes from $\theta=0^{\circ}$ (at the north pole) to $180^{\circ}$
(at the south pole) with a step of $30^{\circ}$, and contains
a convectively unstable layer, surrounded by stable stratified
layers with overshooting convection. 
The height of the convection zone $d$ is chosen as the unit length.

The box rotates around the polar axis from west to east (the angular
velocity vector $\vec{\Om}$ points toward the north pole).
The geometry of the computational domain is
$(x,y,z)\in[0,8d]\in[0,8d]\in[-2d,0]$. This volume is discretized by 
$100\times 100\times 128$ grid points which are uniformly distributed in
each coordinate direction.

The governing equations describing thermal convection in a rotating
stratified medium 
are 
\begin{eqnarray}
\lefteqn{\frac{\partial\rho}{\partial t} = -\nabla\cdot(\rho\vec{u}),}\label{basic1}\\
\lefteqn{\frac{\partial\rho\vec{u}}{\partial t} = -\nabla\cdot(\rho\vec{u}\vec{u})-\nabla P
+\nabla\cdot\pi + \rho\vec{g} - 2\rho\vec{\Om}\times\vec{u},}\label{basic2}\\
\lefteqn{\frac{\partial \rho U}{\partial t} =
-\nabla\cdot(\rho U\vec{u})-P\nabla\cdot\vec{u}+Q_{\rm vis}+
\nabla\cdot(\rho C_{\rm p} \chi\nabla T).}
\label{basic3}
\end{eqnarray}
The notation for physical variables is standard ($U$ thermal energy density, $\pi$ the 
 viscous
stress tensor and $Q_{\rm vis}$ the 
viscous heating term.

The equations~(\ref{basic1})--(\ref{basic3}) are closed through the
ideal gas equation 
$
P=({\cal R}/\mu) \rho T$.
The initial distribution of the physical quantities  represents a
3-layer polytrophic stratification.
We assume all quantities to be periodic in the horizontal 
directions. At the bottom ($z=-2$) and top ($z=0$) of the box
impermeable conditions are imposed for the vertical velocity, while
the horizontal velocities satisfy stress-free boundary conditions.
The temperature and density are fixed at the top of the domain
and a constant heat-flux is injected at the bottom.

The dimensionless parameters ${\rm Ra},\ {\rm Pr}$ and ${\rm Ta}$ are used to control the
simulations. 
In our calculations ${\rm Ra}=3\cdot10^5$ and ${\rm Pr}=0.1$, while
${\rm Ta}\in\{10^5,\ 10^6\}$.

%%%%%%%%%%%%%%%%%%%%%%%%%%%%%%%%%%%%%%%%%%
\section{Rotation-induced anisotropic  turbulence }
\begin{figure*}
\hbox{
\psfig{figure=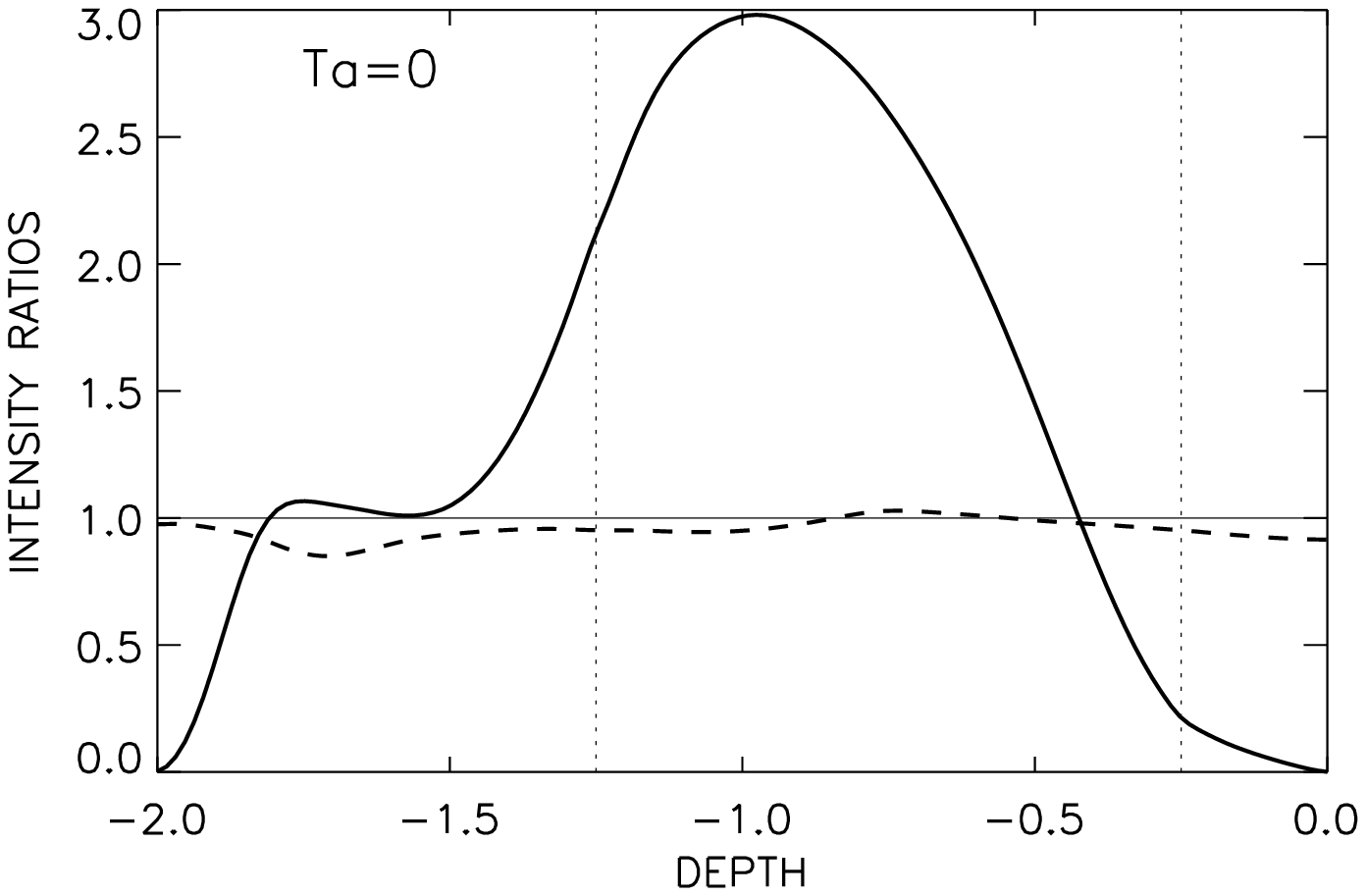,height=5cm,width=6cm}
\psfig{figure=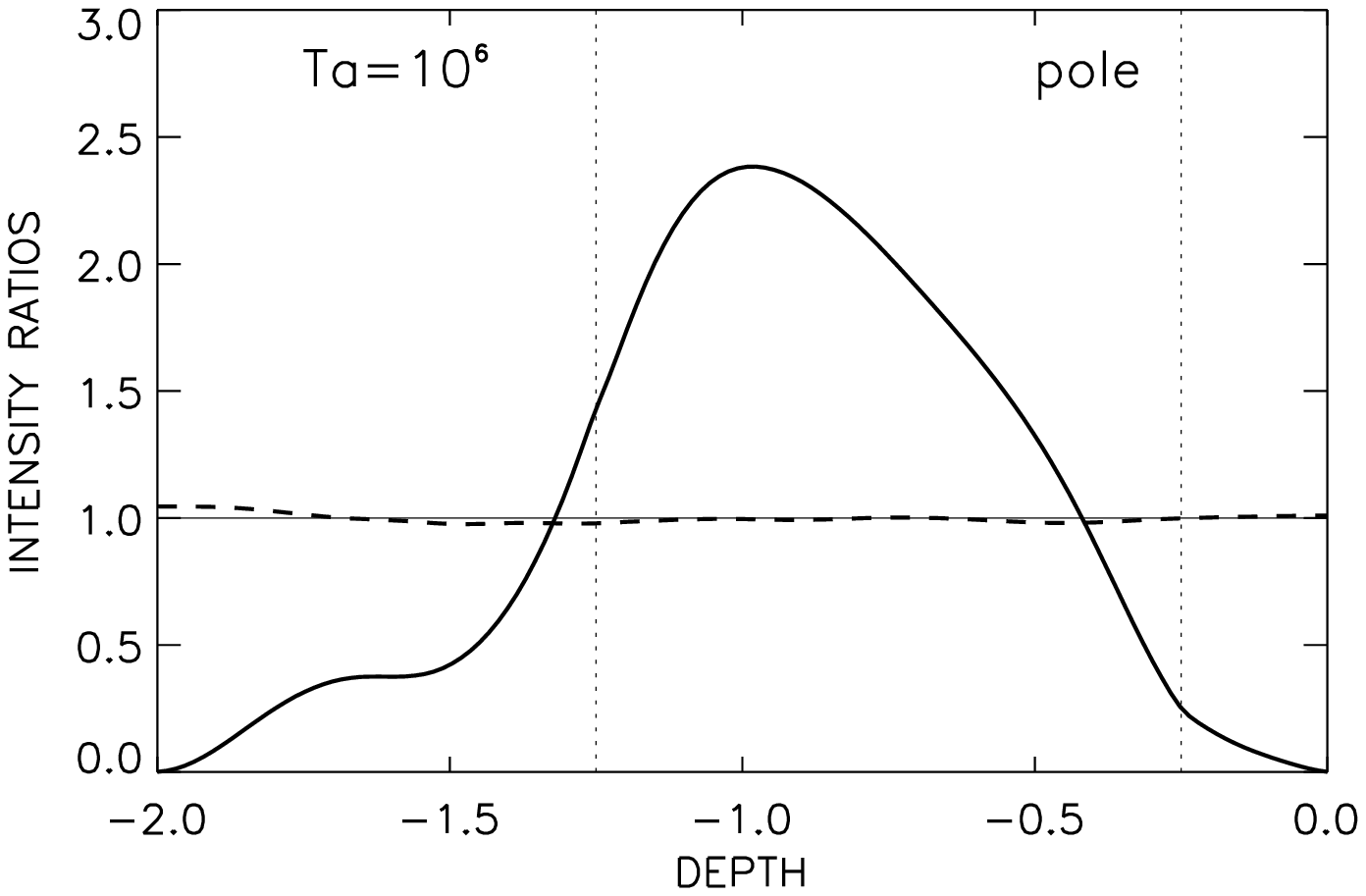,height=5cm,width=6cm}
\psfig{figure=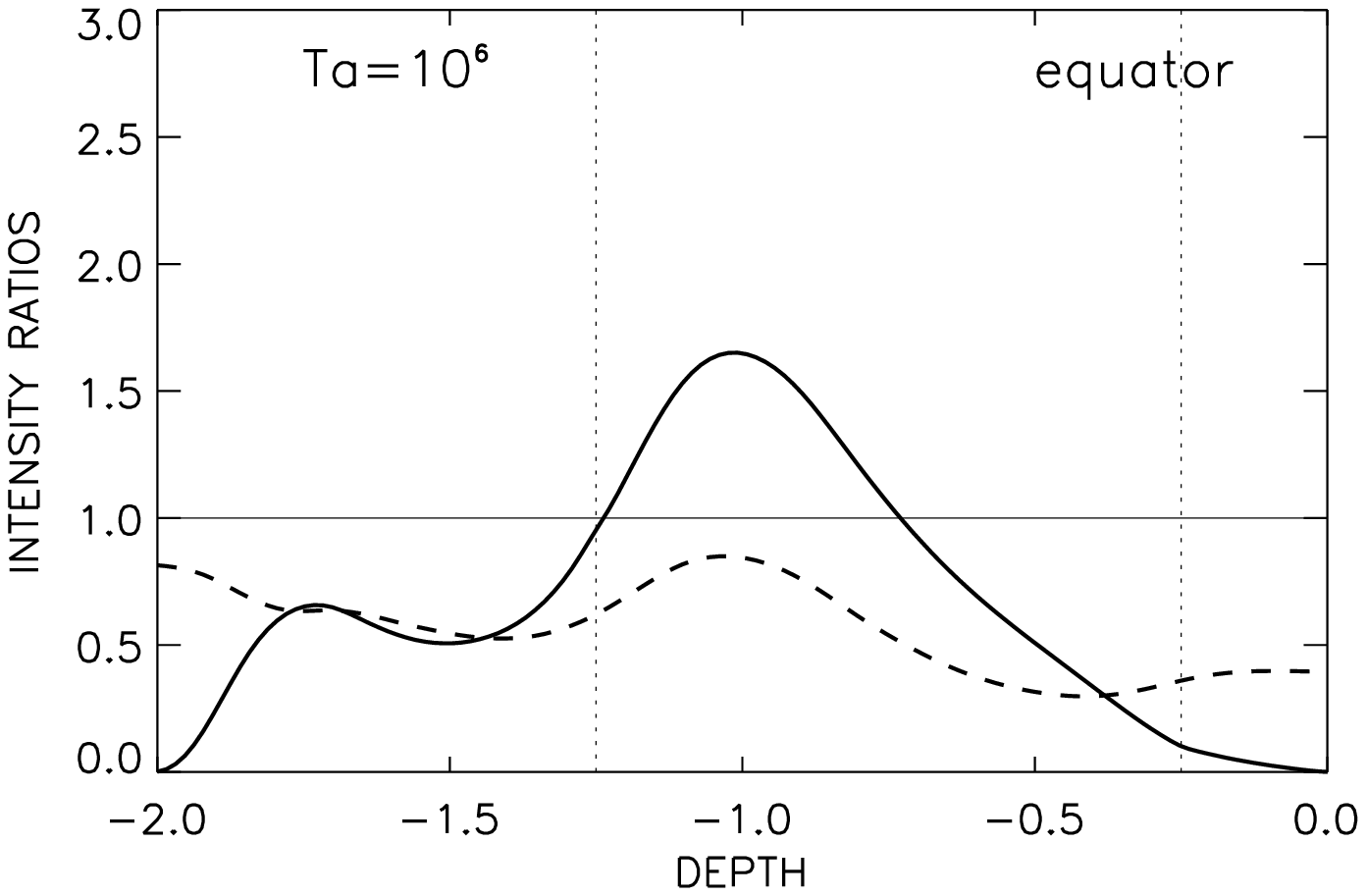,height=5cm,width=6cm}}
\caption{Ratios of the turbulence intensity for the nonrotating  and rotating turbulence fields. Solid: 
$\langle u_r'^2\rangle/\langle u_\phi'^2\rangle$, dashed: 
$\langle u_\theta'^2\rangle/\langle u_\phi'^2\rangle$. {\em Left}: Except the
surface layers the turbulence is strongly vertical-dominated. It is even
isotropic
 in the lower overshoot region. {\em Middle}: 
${\rm Ta}=10^6$, pole. {\em Right}: ${\rm Ta}=10^6$, equator. Under the
influence of rotation only the lower half of the box remains vertical-dominated.
Note that the basic rotation originates a (mild) dominance of the horizontal
motions over the radial motions in the lower overshoot region.
}
\label{anisorot}
\end{figure*}

We start with a discussion of the basic anisotropy between vertical and horizontal turbulence intensities without (Fig. \ref{anisorot}, left) and with rotation (Fig. \ref{anisorot}, middle \& right). 
Without rotation except in the top layer the turbulence is vertically dominated
(Fig.~\ref{anisorot}, left). As it must be, both the horizontal intensities are
equal. The same is true for ${\rm Ta}=10^6$ in the polar region
(Fig.~\ref{anisorot}, middle). For the equator, however, the vertical-horizontal  anisotropy is
more and more reduced but a new anisotropy developes between both the horizontal components (Fig.~\ref{anisorot}, right).

The simulations also provide the anisotropies in the overshoot region. This is in
particular important for the lower overshoot region in relation to the
tachocline discussion. It has been argued that the stability of this zones
against convection changes the turbulence to the horizontal-type (Spiegel \&
Zahn 1992). This is {\em not} observable in Fig.~\ref{anisorot} for the case
without rotation (left).

\begin{figure}
\psfig{figure=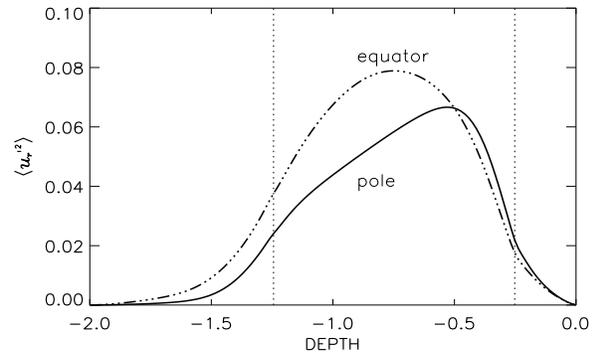,height=5cm,width=8cm}
\caption{The radial turbulence intensity for ${\rm Ta}= 10^6$. Note that except the top layer the turbulence at the equator  exceeds the polar values. Quite a similar result follows from box simulations without any density stratification 
(Giesecke, Ziegler \& R\"udiger 2004).}
\label{radial}
\end{figure}

The behavior of the vertical turbulence intensity is even more important. For
free and anisotropic turbulence we expect for vertically-dominated turbulence
that the polar values exceed the equatorial values (see Appendix). The opposite behavior 
is shown by the simulations. Fig.~\ref{radial} reveals  that in the bulk
of the convection box the equator dominates the poles, i.e.
\beg
\langle u_r'^2\rangle \bigg|_{\rm
eq}>  \langle u_r'^2\rangle \bigg|_{\rm pole} .
\label{urphi}
\ende
This unexpected result (see Eq.~(\ref{upolequat})) has important consequences for the meridional components
of the heat-flux. After Eq.~(\ref{11}) the heat-conductivity tensor is proportionate to the one-point correlation tensor (multiplied with a correlation time) so that it should not be too surprising if also the radial heat-flux $F_r$ would peak at the equator rather than at the poles. The simulations confirm this expectation.
%%%%%%%%%%%%%%%%%%%%%%%%%%%%%%%%%%%%%%%%%%%%%%%%%%%%%%%%%%%%%%%%%%
\section{The poloidal heat-flux}\label{s5}
\begin{figure}
\vbox{
\psfig{figure=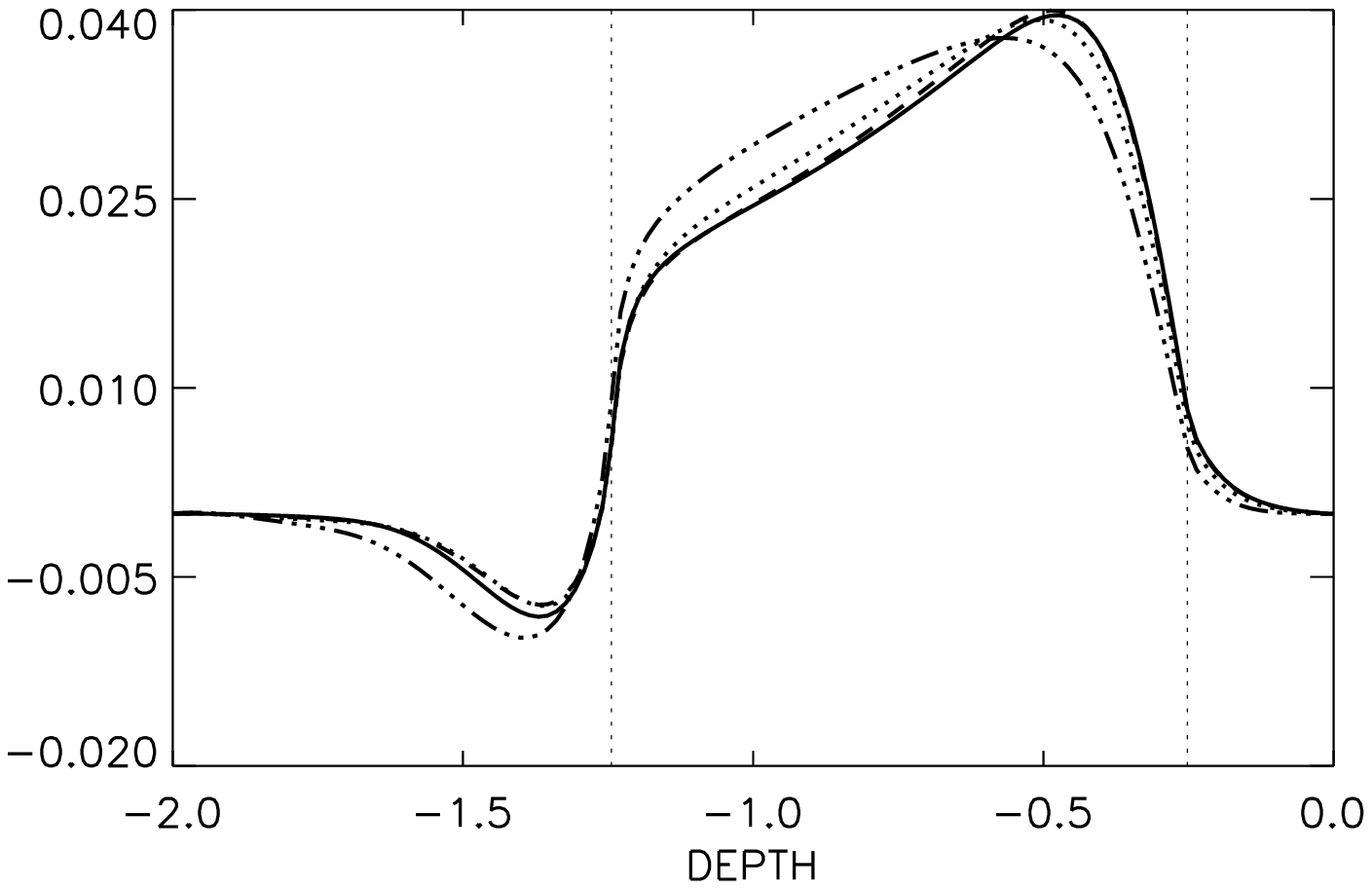,height=5cm,width=8cm}
\psfig{figure=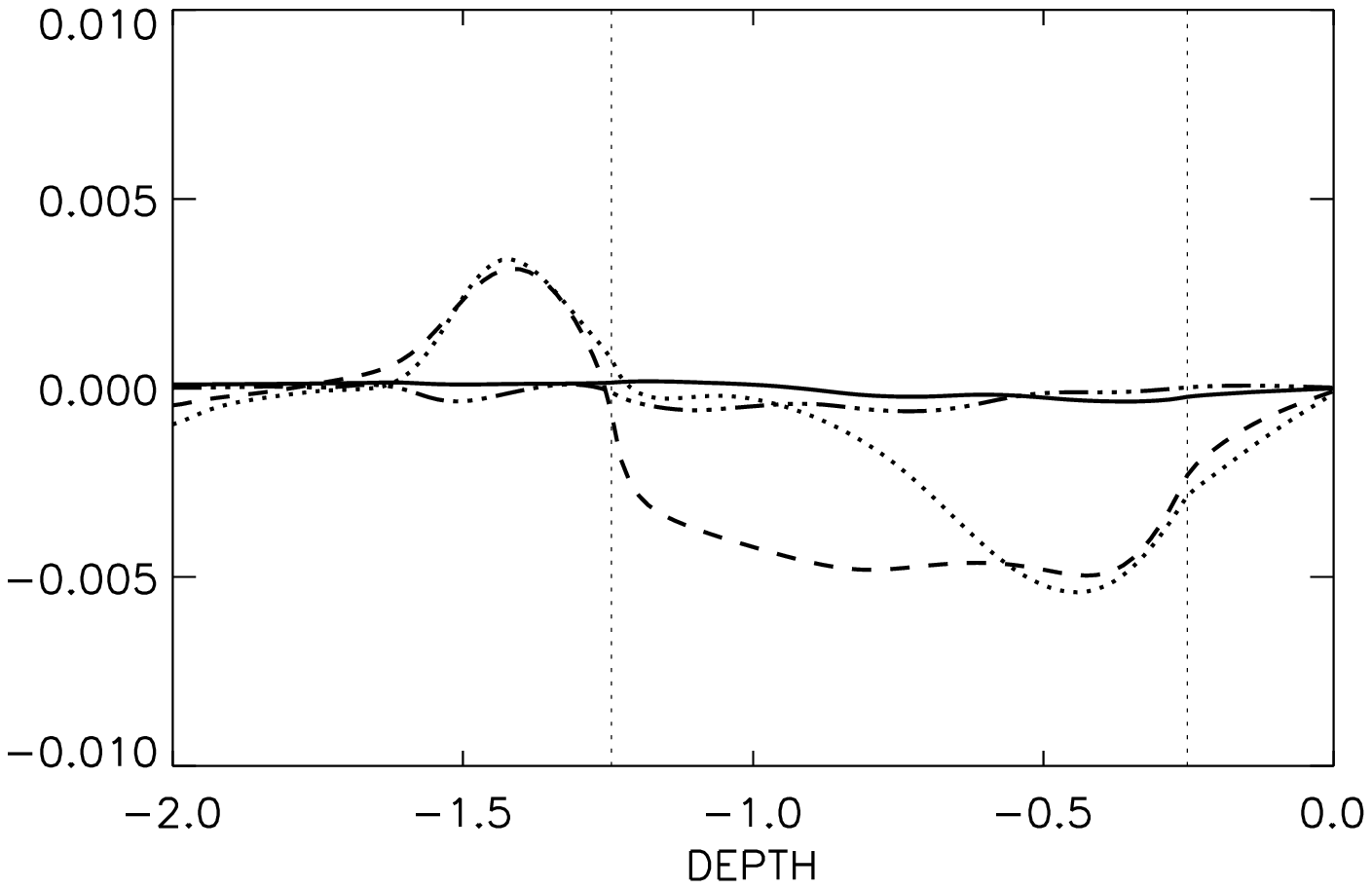,height=5cm,width=8cm}
}
\caption{The correlations $\langle\vec{u}'T'\rangle$ for different colatitudes in the box simulations 
after horizontally and time-averaging vs depth, ${\rm Ta}=10^6$. Top: $\langle u_r'T'\rangle$, bottom: $\langle u_\theta'T'\rangle$. Solid: pole, dashed: 30$^\circ$, dotted: 60$^\circ$, triple-dot-dashed: equator.}
\label{figure01}
\end{figure}

%%%%%%%%%%%%%%%%%%%%%%%%%%%%%%%%%%%%%%%%%%%%%%%%%%%%%%%%%%%%%%%%%%%%%%%
Figure~\ref{figure01} (top) shows the depth-profile of the correlation $\langle
u_r' T'\rangle$ in the box for various latitudes. Due to the rotation the values
differ for poles and equator. The pole-equator difference, however, of the
radial heat-flux depends on the radius. Except the top layer the eddy heat-flux
at the equator exceeds the eddy heat-flux at the poles. In the top layers
where after Fig.~\ref{anisorot} the turbulence is horizontally-dominated the polar
heat-flux dominates the equatorial one. 

This is a characteristic but unexpected
result. It does not contradict, however, the findings of Tilgner \& Busse (1997,
their Fig.~8) in which the latitudinal profiles of the radial heat-flux differ
for differing parameters. For the cases with large Pr  at the inner  boundary 
the heat-flux dominates indeed at the equator; and at the outer boundary the heat-flux dominates indeed at the poles.

Also Rieutord et al. (1994, their Fig. 8a) and K\"apyl\"a, Korpi \& Tuominen (2004, their Fig.~7) found 
similar results. Here we are led to the general conclusion  that a
crossover exists of the pole-equator difference of the radial eddy heat-flux
almost at the same depth where the vertically-dominated turbulence changes to a
horizontally-dominated turbulence. As we have demonstrated with Eq.~(\ref{11}) the
behavior of the radial heat-flux is a direct reflection of the
rotation-influenced radial turbulence intensity $\langle u_r'^2\rangle$. It is
shown in Fig.~\ref{radial} that in the box (except the outermost layer) the $\langle
u_r'^2\rangle$ at the equator exceeds the value at the poles.

A similar crossover does not exist for the latitudinal eddy heat-flux $\langle
u_\theta' T'\rangle$ plotted in Fig.~\ref{figure01} (bottom). This heat-flux
vanishes by definition at the poles and the equator. Between these extrema the
heat flows {\em towards the pole} in the convection zone (and towards the equator
in the lower overshoot region). It is a consequence of the Coriolis force which
is {\em not} involved in the description (\ref{0}). The function $\chi_\parallel$
in (\ref{5.3})$_2$, obviously, is positive-definite as was predicted in
Section~\ref{s21} by a much simpler consideration.

%%%%%%%%%%%%%%%%
%%%%%%%%%%%%%%%%
%%%%%%%%%%%%%%%%
\begin{figure}
\psfig{figure=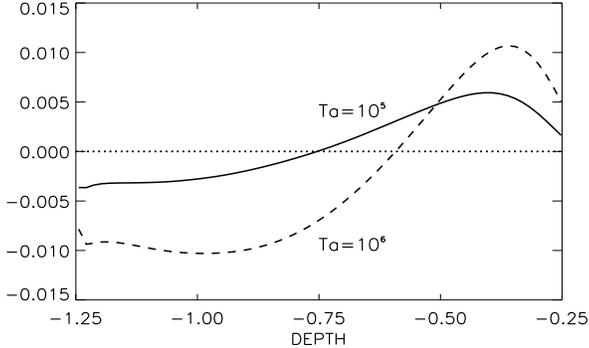,height=5cm,width=8cm}
\caption{The coefficient function $\hat \chi$ in the box for different Taylor
numbers. Note the characteristic zero close to the top of the convective box.}
\label{chihat}
\end{figure}
%%%%%%%%%%%%%%%%
\section{The azimuthal heat-flux}\label{s6}

 Obviously, the last term in (\ref{3}) is linear in $\Om$ in
opposition to the second one which only appears in the second order in
$\Om$. The expression linear in $\Om$ leads to an {\em
azimuthal} component of the heat-flux of
\beg
F_\phi=\rho C_{\rm p}  \chi_{\phi r} \beta_r \Om ,
\label{4}
\ende
where $\beta_r >0$ in convection zones. The sign of
$ \chi_{\phi r}$, therefore, determines the sign of $F_\phi$. One could argue,
however, that for axisymmetric constellations the $F_\phi$ is not important
because of ${\rm div}\,\vec{F}=0$ in that case but i) many stars are not
axisymmetric and ii) the linear-in-$\Om$ term $F_\phi$ should be important
for the comparison of (quasilinear) theory and (nonlinear) simulation. If the
rather simple analytical theory of $F_\phi$ cannot be confirmed by numerical
simulations then the chance is small to understand the higher-order terms in
(\ref{3}) which concern the poloidal components of the vector $\vec{F}$.

The azimuthal heat-flux results from the relation
\beg
\chi_{\phi r}=\sin\theta \Om \tilde\chi.
\label{6}
\ende
The quantity $\tilde \chi$ in Eq.~(\ref{6}) has been computed for a simple
turbulence model subject to a basic rotation but without stratification by
Kitchatinov, Pipin \& R\"udiger (1994). The term was positive-definite but it proved to be
very small. The main reason for its smallness is that the integral kernel runs
with $\omega^2$ so that in the $\tau$-approximation the  term vanishes.

\begin{figure}
\psfig{figure=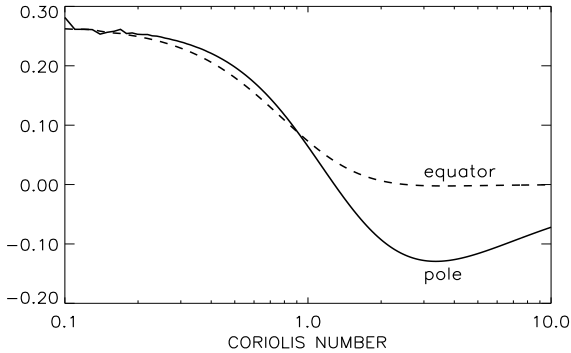,height=5cm,width=8cm}
\caption{The $\Lambda$-effect quantity $V$ as a function of latitude and
Coriolis number (i.e. $\Om$). It is mainly negative between pole and equator.
The turbulence model is highly density-stratified.}
\label{figure1}
\end{figure}

\begin{figure}
\psfig{figure=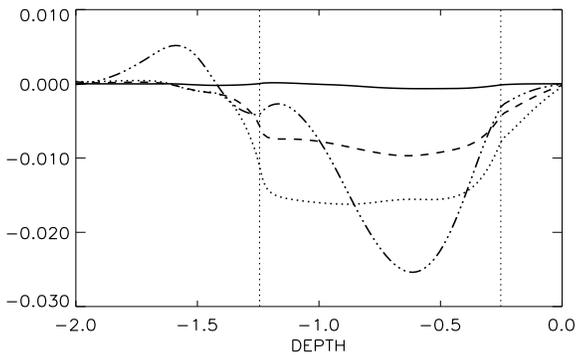,height=5cm,width=8cm}
\caption{The correlation $\langle{u_\phi}'T'\rangle$ for different colatitudes in the box simulations 
after horizontally and time-averaging vs depth, ${\rm Ta}=10^6$.  }
\label{figure2}
\end{figure}

This is not true if anisotropic turbulence is considered. After
(\ref{chiij2}) we have
\beg
\chi_{\phi r}=\frac{1}{2} \tau_{\rm corr} Q_{\phi r}.
\label{12}
\ende
As known the zonal cross correlation $Q_{\phi r}$ only exists for rotating
stratified turbulence and we write 
\beg
Q_{\phi r}= \nu_{\rm T} \Om V \sin\theta 
\label{13}
\ende
with $V$ as the vertical component of the $\Lambda$-effect. Along this way we
arrive at
\beg
\chi_{\phi r}=\frac{\Omst}{4} \nu_{\rm T} V \sin\theta.
\label{14}
\ende
The zonal heat-flux (\ref{4}) should thus be positive (negative) for positive
(negative) $V$. A very similar relation has been formulated by  Gough (1978).
Obviously, the $F_\phi$ and the radial $\Lambda$-effect have the same sign. In
Fig.~\ref{figure1} the radial $\Lambda$-effect quantity $V$ is given obtained by
Kitchatinov \& R\"udiger (1993) within the quasilinear $\tau$-approximation.
Generally, the function $V$ is negative which result has recently been confirmed
numerically by Chan (2001) and K\"apyl\"a, Korpi \& Tuominen (2004). Hence, the
$F_\phi$ should be negative and this is also the result of the simulations 
shown in Fig.~\ref{figure2}. Positive (negative) fluctuations $u_\phi'$ are thus
be correlated with negative (positive) temperature fluctuations which should be
observable.

%%%%%%%%%%%%%%%%

%%%%%%%%%%%%%%%%%%%%%%%%%%%%%%%%%%%%%%
\section{Rotation-law consequences }\label{s7}
As mentioned in Section~\ref{s1} the rotation laws which  have been computed so far
(Kitchatinov  \& \R\ 1999;  K\"uker \& Stix 2001) were obtained with a radial heat-flux which peaks at the poles as the result
of the basic rotation. The simulations do not confirm this latitudinal profile. On
the other hand, the rotation laws are computed with a latitudinal heat-flux
which goes to the poles and this is confirmed by the simulations. We have thus
to check the consequences of the new situation for the rotation law theory.

\begin{figure}
\vbox{
\psfig{figure=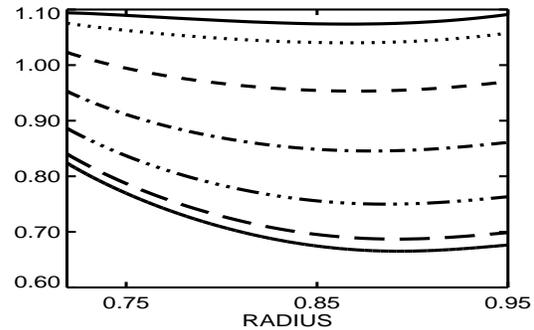,height=5.5cm,width=8cm}
\psfig{figure=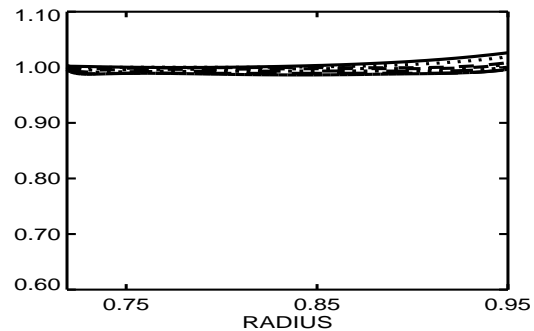,height=5.5cm,width=8cm}
}
\caption{The rotation law in the solar convection zone without the influence of the rotation-modified 
heat-flux without ({\em top}) and with ({\em bottom}) meridional circulation.}
\label{figrotlaw1}
\end{figure}

As a demonstration of the complex character of the resulting mean-field rotation
laws in Fig.~\ref{figrotlaw1} (top) the rotation law is given as due to the 
$\Lambda$-effect alone
(details given by  K\"uker \& Stix 2001) without both  meridional flow and  rotation-induced eddy heat-flux. The rotation
profile  well complies with the observations (see K\"uker, R\"udiger \&
Kitchatinov 1993).

The inclusion of the meridional flow (Fig.~\ref{figrotlaw1}, bottom) drastically changes the
situation. The resulting poleward surface flow strongly reduces the equator-pole
difference and according to the Taylor-Proudman theorem the $\Om$-isolines 
become parallel to the rotation axis. In this case one  finds a slight  superrotation beneath
 the equator and a rather uniform  angular velocity beneath the pole  -- in
great contrast to the observations.

\begin{figure}
\vbox{
\psfig{figure=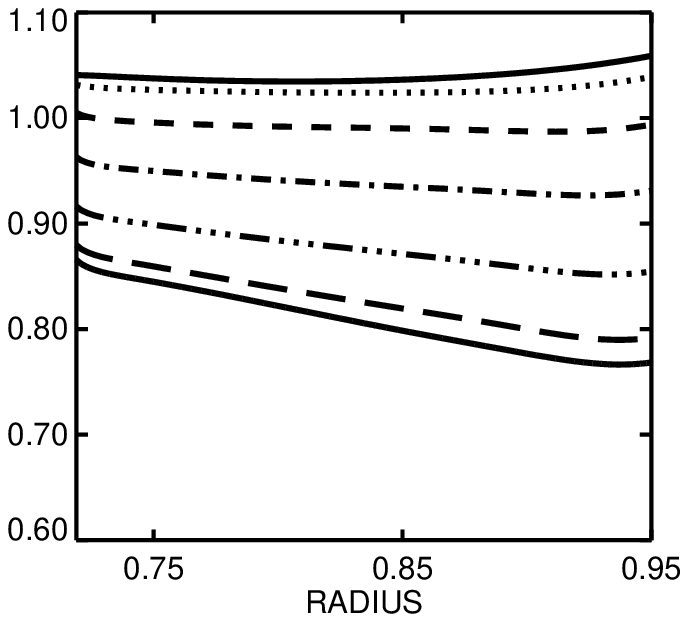,height=5.5cm,width=8cm}
\psfig{figure=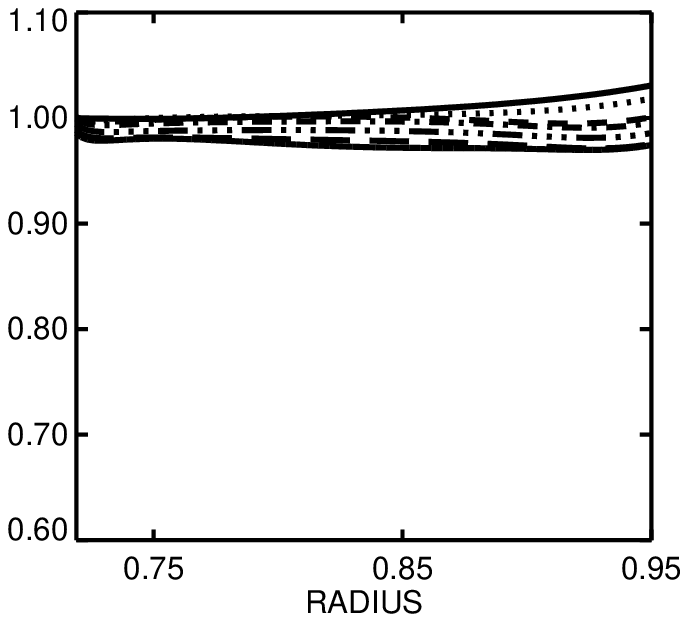,height=5.5cm,width=8cm}
}
\caption{The same as in Fig.~\ref{figrotlaw1} but with the full 
rotation-modified heat-flux ({\em top}) and  $\chi_{\theta r}=0$ 
(see Eq.~(\ref{5.3})$_2$ {\em bottom}). Obviously, the rotation-induced off-diagonal component $\chi_{\theta r}$ produces the main effects.}
\label{figrotlaw2}
\end{figure}

The situation changes with the rotation-induced eddy heat-flux included. The
poles become warm and a circulation developes towards the equator. Hence, the
total meridional flow becomes rather slow and again we have the situation of Fig.~\ref{figrotlaw1}
(top). This is the solution of the `Taylor number puzzle'
which consists in the existence of two opposite directions of the meridional
circulation (Fig.~\ref{figrotlaw2}, top). In Fig.~\ref{figrotlaw2} (bottom) the numerical experiment
with (artificially) $\chi_{\theta r}=0$ reveals that it is the $\chi_{\theta r}$-effect which
solves the Taylor number puzzle rather than the latitude-profile of $\chi_{rr}$. 

\section{Discussion}
For rotating free anisotropic turbulence Eq.~(\ref{upolequat}) provides a dominance of the
vertical turbulence intensity at the poles. Our box simulations reveal the
opposite behavior. As shown in Fig.~\ref{radial} the turbulence at the equator
dominates. This is {\em not} a consequence of the density stratification which
was not involved in the evaluation of the correlation tensor for rotating free
turbulence as given in the Appendix. Giesecke, Ziegler \& \R\ (2004) considered boxes
without any density stratification but with the same result. One needs global
simulations in order to find out whether or not the limited size of the boxes is
responsible for the unexpected effect.

An immediate consequence of this result concerns the
latitude-dependence of the vertical turbulent heat-transport which is given in
Fig.~\ref{figure01} (top). In the main bulk of the box the heat-flux at the equator
exceeds the vertical heat-flux at the poles\footnote{In the top layer of the 
box
where after Fig.~\ref{radial} the polar value of $\langle u_r'^2\rangle$ dominates the
equatorial one also the polar value of the heat-flux dominates its equatorial
value}. For free
turbulence the quasilinear theory on the basis of the $\tau$-approximation leads
to the opposite behavior. There is possibly a simple explanation of the
differences for the latitudinal profile of $\langle u_r'^2\rangle$ for free
turbulence and for box simulations. For rapid rotation the Taylor-Proudman theorem
strongly damps the $u_r$-component of the turbulence at the poles due to the
boundary condition $u_r'=0$ at the top and bottom of the box.

This argument hardly holds for the cross-correlations $\langle u_\theta'
u_r'\rangle$ and $\langle u_\phi' u_r'\rangle$ which identically vanish at the
poles (see Eqs.~(\ref{a15}) and (\ref{a16})). The results of the simulations for the
eddy heat-flux components $F_\theta$ and $F_\phi$ do fully comply with the
analytical quasilinear expressions. Both correlations are negative at the northern
hemisphere\footnote{$F_\phi$ is also negative at the southern hemisphere}. The latitudinal heat transport is therefore always polewards
and the azimuthal heat-flux is always westwards. The results of K\"apyl\"a, Korpi \& Tuominen
(2004) for the latitudinal heat-transport are here less coherent.

The correlation  $\langle u_\phi' T'\rangle$ is due to the rotational influence and it represents the radial angular momentum transport by the turbulence. It should be observable at the solar surface. 
Convective-originated temperature fluctuations like those of the mesogranulation are expected to be anticorrelated with their local angular velocity fluctuations (see Duvall \& Gizon 2000) . 

For axisymmetric stars, of course,  the longitudinal heat-flux is not important
because of $\partial F_\phi/\partial \phi=0$ but this might not to be true for 
close binaries or also for single stars with distinct nonaxisymmetric surface 
structures like the (flip-flop) FK Com stars. 

More clear is the meaning of the latitudinal heat-flux component $F_\theta$. It is always polewards directed leading to the formation of warm polar caps -- not in agreement with the results of Rieutord et al. (1994). We have shown with the simulations presented by   the Figs. \ref{figrotlaw1} and \ref{figrotlaw2} how important this effect for the explanation of the solar/stellar differential rotation is. The meridional flow driven by the warm poles breaks the Taylor-Proudman theorem which would otherwise produce cylindrical isolines of the angular velocity which are not found  by the the inversions of the 
helioseismology though. As we have also shown, the role of the radial eddy heat-flux $F_r$ is here of minor importance.
%%%%%%%%%%%%%%%%%%%%%%%%%%%%%%%%%%%%%%
\begin{acknowledgements}
Axel Brandenburg (Kopenhagen) is acknowledged for many discussions about the  subject of this paper. LLK is grateful to AIP for its hospitality and the visitor support.
\end{acknowledgements}

%%%%%%%%%%%%%%%%
%%%%%%%%%%%%%%%%%%%%%%%%%%%%%%%%%%%%%%%%%%%%%%%%%%%%%%%%%%%%%%%%%%%%%%%%
\appendix
%%%%%%%%%%%%%%%%%%%%%%%%%%%%%%%%%%%%%%%%%%%%%%%%%%%%%%%%%%%%%%%%%%%%%%%%
\section{Rotating anisotropic turbulence}\label{SOCA}
%%%%%%%%%%%%%%%%%%%%%%%%%%%%%%%%%%%%%%%%%%%%%%%%%%%%%%%%%%%%%%%%%%%%%%%%
The usual approach to rotating turbulences is to \emph{prescribe} the
turbulence for nonrotating fluid and then to \emph{derive} the influence of
rotation on the given original turbulence.
As the first step in this procedure, we consider an incompressible
fluctuating velocity field, $\vec u^{(0)}$, in a nonrotating
fluid. The statistical properties of the original turbulence are given by its
spectral tensor,
\begin{eqnarray}
 \hat{Q}^{(0)}_{ij} &=& {E(k,\omega )\over 16\pi k^2} K_{ij}({\vec k})
 \nonumber \\
 &+&
 {E_1(k,\omega )\over 16\pi k^4}\left(K_{ij}({\vec G})K_{mn}({\vec G}) -
 K_{im}({\vec G})K_{jn}({\vec G})\right)k_m k_n ,
 \label{a1}
\end{eqnarray}
where $\vec G$ is the radial unit vector, $\vec k$ is the wave vector,
$\omega$ is frequency, and $K_{ij}$ is the projection tensor,
\begin{equation}
K_{ij}({\vec k}) = \delta_{ij} - k_i k_j/k^2 .
 \label{a2}
\end{equation}
Without rotation, the only preferred direction can be the radial one.
Equation (\ref{a1}) allows for the radial anisotropy. Characteristic
velocities are given by 
\begin{eqnarray}
\lefteqn{ \langle u_r^{(0)2}\rangle = {1\over 3}\int\limits_0^\infty E(k,\omega )\ {\rm
d}k{\rm d}\omega,}
 \nonumber \\
\lefteqn{ \langle u_{\rm h}^{(0)2}\rangle - 2\langle u_r^{(0)2}\rangle
 = {1\over 3}\int\limits_0^\infty E_1(k,\omega )\
 {\rm d}k{\rm d}\omega,}
 \label{a3}
\end{eqnarray}
where $u_{\rm h}^{(0)}$ is the horizontal velocity, $\langle u_{\rm h}^{(0)2}\rangle =
\langle u_\phi^{(0)2}\rangle +
\langle u_\theta^{(0)2}\rangle$.
Spectral functions must satisfy $E_1 \geq -E$. A dimensionless anisotropy parameter,
$a$, is introduced
\begin{equation}
 {a} = \left(\langle u^{(0)2}_{\rm h}\rangle
 - 2\langle u^{(0)2}_r\rangle\right)/\langle u^{(0)2}_r\rangle ,
 \label{a4}
\end{equation}
with $a \geq -1$ (Bochner's theorem). Negative $a$ means anisotropy of radial type with
$\langle u^{(0)2}_\phi\rangle  = \langle u^{(0)2}_\theta\rangle < \langle
u^{(0)2}_r\rangle$
while positive $a$ means predominance of horizontal motions.

The rotational influence is involved via the tensor 
\begin{equation}
 D_{ij}\left({\vec\Om}\right) = {\delta_{ij} + {\left(2\bm {k}\cdot\bm{\Om}\right)
 \over k^2\left(- {\rm i}\omega+\nu_{\rm t}k^2\right)} \epsilon_{ijp} k_p
 \over 1 + {\left( 2\bm{k}\cdot\bm{\Om}\right)^2\over
 k^2\left(- {\rm i}\omega+\nu_{\rm t}k^2\right)^2}} ,
 \label{a5}
\end{equation}
(R\"udiger 1989), which expresses the rotating turbulence in terms of the original one by the
linear relation 
\begin{equation}
 \hat{Q}_{ij}\left( {\vec k},\omega\right) = D_{im}\left( {\vec\Om}\right)
 D_{jn}^*\left({\vec\Om}\right)\hat{Q}^{(0)}_{mn}\left( {\vec k},\omega\right) ,
 \label{a6}
\end{equation}
where the star here means complex conjugate and $\nu_{\rm t}$ is the effective viscosity by microscale turbulence.

With (\ref{a6}), the parameters of the rotating turbulence can be derived.
We make, however, further simplifications
by introducing the mixing-length approximation via a
special form of the spectral functions 
\beg
E\left( k,\omega\right) = 6\cdot\langle u^{(0)2}_r\rangle 
\delta\left( k - \ell_{\rm corr}^{-1}\right)
  \delta\left(\omega\right) ,\quad \quad \ E_1 = a E ,
 \label{a7}
 \ende
 with
$\ell_{\rm corr}^2/\nu_{\rm t} = \tau_{\rm corr}$ 
(Kitchatinov 1991), where $\ell_{\rm corr}$ and $\tau_{\rm corr}$ are the mixing length and time respectively. An advantage
of (\ref{a7}) is that the effects of rotation can be expressed
in terms of relatively simple parameters like the Coriolis number.
The tensor (\ref{a5}) now reads
\begin{equation}
 D_{ij}\left({\vec\Om}\right) = {\delta_{ij} + \cos\theta \Omst\ \epsilon_{ijp}
 k_p/k\over 1 +  \cos^2\theta {\Omst}^2},
 \label{a9}
\end{equation}
where $\theta$ is
the angle between the angular velocity and the wave vector.

The relation~(\ref{a6}) can now be integrated over the wave number space to find the
one-point-correlation tensor. It reads
\begin{eqnarray}
 Q_{ij} &=& \langle u_r^{(0)2}\rangle \left\{\left(\phi\left(\Omst\right)\delta_{ij}\
 +\ \phi_\|\left(\Omst\right){\Om_i\Om_j\over\Om^2}\right)+\right.
 \nonumber \\
 &+&a \bigg(\phi'\left(\Omst\right)\delta_{ij}\
 +\ \phi_1\left(\Omst\right){\left({\vec\Om}\cdot{\vec G}\right)^2
 \over\Om^4}\Om_i\Om_j +
 \nonumber \\
 &+& \phi_2\left(\Omst\right)G_iG_j\ -\ \phi_3\left(\Omst\right)
 {\left({\vec\Om}\cdot{\vec G}\right)\over\Om^2}\left(\Om_iG_j\ +\
 \Om_jG_i\right)+
 \nonumber \\
 &+& \left(\phi'_\|\left(\Omst\right)/2 + \phi_3\left(\Omst\right)\right)
 \left(\left({\vec\Om}\cdot{\vec G}\right)^2\delta_{ij}\ +\ \Om_i\Om_j\right)
 /\Om^2 \bigg)+
 \nonumber \\
 &+&a I_0\left(\Omst\right)\left(G_i\epsilon_{jmp} +
 G_j\epsilon_{imp}\right) {\Om_m\over\Om} G_p+
 \nonumber \\
 &+& \left. a I_1\left(\Omst\right)
 {\left({\vec\Om}\cdot{\vec G}\right)\over\Om^3}
 \left(\Om_i\epsilon_{jmp} + \Om_j\epsilon_{imp}\right)
 \Om_mG_p \right\}
 \label{a11}
 \end{eqnarray}
  for rotating anisotropic turbulence.
  For slow rotation 
\begin{equation}
 \phi \simeq 1,\ \phi' \simeq 1/2,\ \phi_2 \simeq -1/2,\ I_0 \simeq 2\Omst/5,
 \label{a12}
\end{equation}
 ($\Omst\ll 1$, other functions are of second or of higher order in
 $\Omst$), and for
 fast rotation ($\Omst \gg 1$)
\begin{eqnarray}
\lefteqn{ \phi = \phi_\| \simeq {3\pi\over 8 \Omst}, \ \phi' \simeq {3\pi\over 64\Omst},
 \ \phi'_\| \simeq \phi_2 \simeq \phi_3 \simeq {3\pi\over 32\Omst},}
 \nonumber \\
\lefteqn{ \phi_1 \simeq {9\pi\over 64\Omst}, \ I_0 \simeq {3\pi\over 16{\Omst}^2},
 \ I_1 \simeq {9\pi\over 16{\Omst}^2} }
 \label{a13}
\end{eqnarray}
($\Omst \gg 1$). Full expressions for the functions are too bulky to reproduce here. Some of them
can be found elsewhere (Kitchatinov, Pipin \& R\"udiger 1994; Kitchatinov 2004).

For the off-diagonal component, $Q_{\theta r}$ yields
\begin{eqnarray}
\lefteqn{  Q_{\theta r} = - \sin\theta\cos\theta\ \langle u_r^{(0)2}\rangle\big(
  \phi_\|\left(\Omst\right)}
  \nonumber \\
  && \quad \quad \quad \quad \quad + a\left(
  \phi'_\|\left(\Omst\right)/2 +
  \cos^2\theta\ \phi_1\left(\Omst\right) \right)\big).
  \label{a15}
\end{eqnarray}
$Q_{\theta r}$ is thus always negative.

The sign of the off-diagonal component $Q_{\phi r}$ of the correlation 
tensor is only controlled by the turbulence anisotropy, i.e.
\begin{equation}
  Q_{\phi r} = \langle u_r^{(0)2}\rangle a \sin\theta\left( I_0\left(\Omst\right)\
  +\ \cos^2\theta\ I_1\left(\Omst\right)\right) .
  \label{a16}
\end{equation}
$Q_{\phi r}$ is thus
negative for radial-type of anisotropy ($a < 0$). It vanishes for isotropic turbulence ($a=0$) and is one of the known components of the $\Lambda$-effect.
 
With the mixing-length approximation (\ref{a7}), the relation between the
turbulent thermal conductivity and the velocity correlation tensor becomes
\begin{equation}
 \chi_{ij} =\ \tau_{\rm corr} Q_{ij} =\ \tau_{\rm corr}\int \hat{Q}_{ij} \left( {\vec k},
 \omega\right)\ {\rm d}{\vec k}{\rm d}\omega .
 \label{a10}
\end{equation}
Note the factor 2 by which this relation differs from (\ref{chiij2}) which only holds in the limit $\chi \to 0$.
%%%%%%%%%%%%%%%%%%%%%%%%%%%%%%%%%%%%%%%%%%%%%%%%%%%%%%%%%%%%%%%%%%%%%%%

\end{document}